# A 5 GHz LNA for a Radio-Astronomy Experiment

Miguel Bergano
Instituto de Telecomunicações
Aveiro, Portugal
jbergano@av.it.pt

Luís Cupido
Instituto de Plasmas e Fusão Nuclear
Instituto Superior Técnico
Lisboa, Portugal

Armando Rocha
Instituto de Telecomunicações/DETI
Universidade de Aveiro
Aveiro, Portugal

Domingos Barbosa
Instituto de Telecomunicações
Aveiro, Portugal

***Abstract*—The paper describes the project, implementation and test of a C-band (5GHz) Low Noise Amplifier (LNA) using new low noise Pseudomorphic High Electron Mobility Transistors (pHEMTS) from Avago. The amplifier was developed to be used as a cost effective solution in a receiver chain for Galactic Emission Mapping (GEM-P) project in Portugal with the objective of finding affordable solutions not requiring strong cryogenic operation, as is the case of massive projects like the Square Kilometer Array (SKA), in Earth Sensing projects and other niches like microwave reflectometry. The particular application and amplifier requirements are first introduced. Several commercially available low noise devices were selected and the noise performance simulated. An ultra-low noise pHEMT was used for an implementation that achieved a Noise Figure of 0.6 dB with 13 dB gain at 5 GHz. The design, simulation and measured results of the prototype are presented and discussed.**

***Keywords – Instrumentation, GaAs, high electron-mobility transistor (HEMT), Low Noise Amplifiers (LNA), noise temperature, Radioastronomy***

## I. INTRODUCTION

The equivalent noise temperature of a receiver is determined by the noise performance of the components after the antenna. The LNA, whose performance results from a compromise between the gain and noise budgets, often determines the receiver noise temperature. In our case, the GEM-P Project in Portugal, as described in [1, 2, 3, 4] requires a high sensitivity receiver capable of measuring the polarized sky noise at C-band with a sensitivity of about 0.2 mK. The GEM-P group, that intends to map the Northern Hemisphere, developed a dual channel super heterodyne receiver with a digital backend (Fig. 1) to measure the Stokes parameters of the galactic radiation in a 200 MHz bandwidth centred around 4.9 GHz.

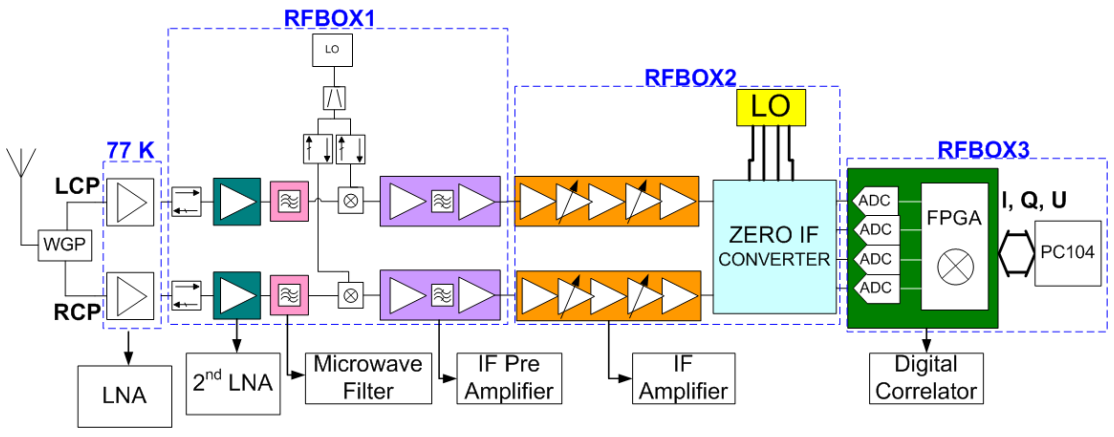

Figure 1. Polarimeter Block Diagram

A 9 meter Cassegrain antenna equipped with an orthomode transducer for circular polarization will feed the two receiver channels. The receivers were developed to deliver a signal, with an adequate amplitude and bandwidth, to two backend digitizers [3]. Each receiver chain uses a first cryogenic LNA with InP based HEMTS (Model LNF-LNC4_8A, Chalmers, Sweden) with 32 dB gain and operating at a physical temperature of 77K. A second RF stage at room temperature and a down conversion to an IF at 600MHz feeds a four channel digital correlator implemented in FPGA (field programmable gate array) [4].

## II. RECEIVER OVERVIEW

The sensitivity of a given radiometer can be described by the radiometer equation that originates from the standard deviation of the power fluctuations of a band limited noise signal:

$$\Delta T = \frac{T_A + T_N}{\sqrt{B \cdot \tau}} \qquad (1)$$

where $T_A$ is the antenna temperature and $T_N$ the system noise temperature, B the detection bandwidth and $\tau$ the integration time. Since receivers are mostly cascaded amplifiers, the LNAs define most of the receiver equivalent noise temperature and, by consequence, the integration time necessary for cosmic signal detection. For feasible system sensitivity we already require many days of integration time so optimizing receiver noise performance is mandatory in order to keep this long integration time to a minimum.

After a power budget and noise temperature analysis (using the cascaded noise Friis equation) a second RF amplifier ($2^{nd}$ LNA in Fig. 1) proves to be useful to improve the receiver performance. The noise factor of a three stage cascade is given by:

$$F = F_1 + \frac{F_2 - 1}{G_1} + \frac{F_3 - 1}{G_1 G_2} \qquad (2)$$

where $F_n$ are the noise factors and $G_n$ the gain of each stage. Assuming, respectively, the F factors to be 1.009, 1.122, 22.387 and the gains 36, 13 and 20 dB we arrive to the values $F_{WITH2^{o}LNA}$=1.0093 and $F_{without2^{o}LNA}$=1.0144. In terms of equivalent noise temperature it corresponds to $T_{EQWITH2^{o}LNA}=2.80$ and $T_{EQwithout2^{o}LNA}=4.32$ K.

This seems a rather marginal improvement but from (1) it can be shown easily that a relative reduction of the receiver noise temperature means twice a relative reduction of the measurement campaign duration. Reducing operating costs and having early results is quite important in any project.

An LNA based on a single transistor would be enough to comply with the requirements. This second stage should be designed to operate at room temperature at the best achievable NF. InP based HEMTS devices as they excel at cryogenic temperatures [10],, but at room temperatures they have a comparable performance to some of the most recent commercially available ultra-low noise pHEMTS that use InGaAs technology. Several of these commercial chips were analysed as candidates for a cost effective LNA. A design with a modern InGaAs pHEMT may provide a very attractive cost/performance relationship at room temperature and may replace second stages previously operating at cryogenic temperatures (77 K) or even provide first stages for less demanding applications.

The noise budget, that as a direct impact on the receiver sensitivity receiver as referred previously, is depicted in TABLE I whose results do not consider cable losses.

TABLE I. RADIOMETER NOISE BUDGET

| | LNA | 2º LNA | FILTER+MIX+IFAMPS |
|---|---|---|---|
| $T_{EQ}$ (K) | 2.7 | 36.6 | 6426.2 |
| NF (dB) | 0.039 | 0.5 | 13.5 |
| GAIN (dB) | 36.0 | 13.0 | 20.0 |
| LINEAR GAIN | 3981.0 | 20.0 | 100.0 |

## III. LOW NOISE AMPLIFIER IMPLEMENTATION

An amplifier is commonly defined by two main characteristics: Gain and noise figure. To have a best compromise between the two characteristics a lossless matching network on both sides of the transistor (Fig. 1) must be designed to transform the input and output impedance $Z_0$ to the source and load impedance $Z_S$ and $Z_L$ required in the design specifications. The most useful gain definition for amplifier design is the transducer power gain, which takes into account both source and load mismatch. Thus from [5], we have separate effective gain contributions: input (source) matching network, the transistor itself and the output (load) matching network as shown in Fig. 2.

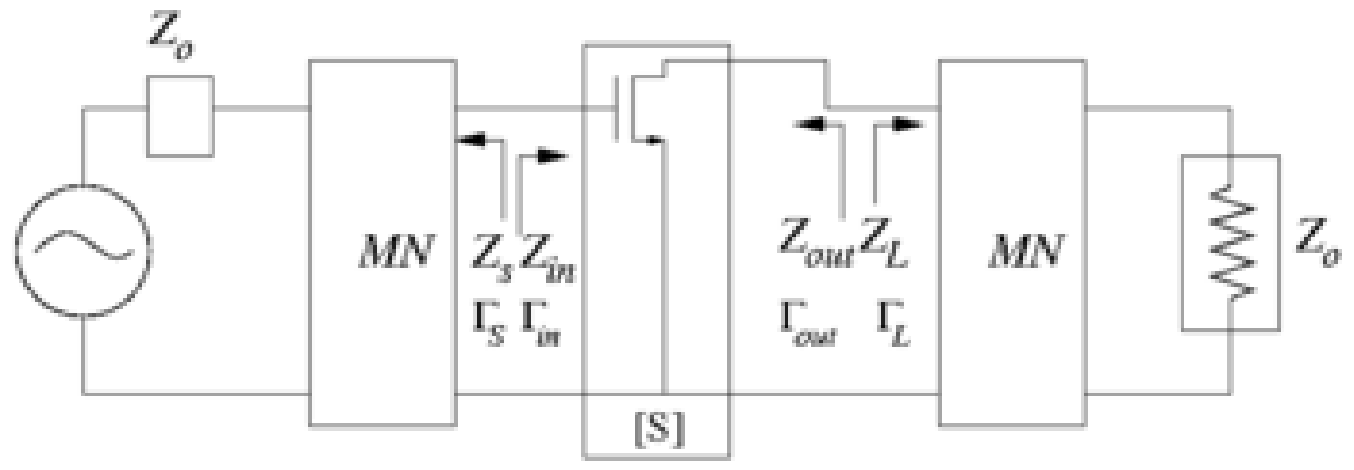

Figure 2. Input/output amplifier matching networks (MN)

The amplifier gain is given by the equation

$$G = \frac{1-|\Gamma_s|^2}{|1-S_{11}\Gamma_s|^2}|S_{21}|^2\frac{1-|\Gamma_L|^2}{|1-S_{22}\Gamma_L|^2} \quad (3)$$

where $S_{11}$, $S_{22}$ and $S_{21}$ are the device scattering parameters, $\Gamma_S$ and $\Gamma_L$ are the reflection coefficients seen from the device towards the matching source and load networks. For maximum gain the source and load impedances ($Z_s$ and $Z_L$) must be complex conjugates of those seen into the device ($Z_{in}$ and $Z_{out}$).

The optimum source impedance to achieve the better NF is not the same as the one that achieves the maximum gain. A compromise between both parameters must be found and the usual method is to plot noise and gain circles in the same Smith chart. The gain circles depend on the S parameters. The noise circles are plotted using the transistor parameters $NF_{MIN}$, $R_N$ and $\Gamma_{OPT}$, i.e. minimum transistor noise figure, the equivalent noise resistance and optimum source reflection coefficient, respectively. Several commercial low noise PHEMT transistors and the newly proposed Heterojunction Bipolar Transistor (HBT) were selected and tested to get the best noise performance. A list of the selected commercial PHEMT and HBT transistors is presented in TABLE II.

TABLE II. COMMERCIAL PHEMT CHARACTERISTICS

| DEVICE | COMPANY | NF (dB) | GAIN (dB) | @FREQ (GHZ) |
|---|---|---|---|---|
| MGF4941 | MITSUBISHI | 0.40 | 13.5 | 12.0 |
| MGFC4419 | MITSUBISHI | 0.35 | 13.5 | 12.0 |
| MGF4931 | MITSUBISHI | 0.50 | 12.0 | 12.0 |
| ATF-36077 | AGILENT | 0.50 | 12.0 | 12.0 |
| NE3511S02 | MITSUBISHI | 0.60 | 11.5 | 12.0 |
| FHX13X | NEC | 0.30 | 13.5 | 12.0 |

A first approach to the LNA design was developed using the manufacturer scattering and noise parameters by plotting the noise and gain circles and all the devices were used in circuit simulations. The software tool used for was the Advanced Design System 2009 from Agilent (ADS) [6].

### A. *Matching Networks*

Using the Smith chart source impedance must be selected as close as possible to the optimum one without neglecting the gain (Fig. 3). As the amplifier main function is to reduce the noise contribution of the devices ahead, the gain here is not at premium. However for the lowest noise figure (NF=0,3dB) the gain is 12dB. The shaded zone in the Smith Chart indicates the possible source impedances of the transistor. The selected value was $Z_S=34+j66\ \Omega$.

The load impedance $Z_L$ seen by the transistor should be the complex conjugate of $Z_{OUT}$. Since the device is not unilateral $Z_{OUT}$ is a function of $Z_S$.

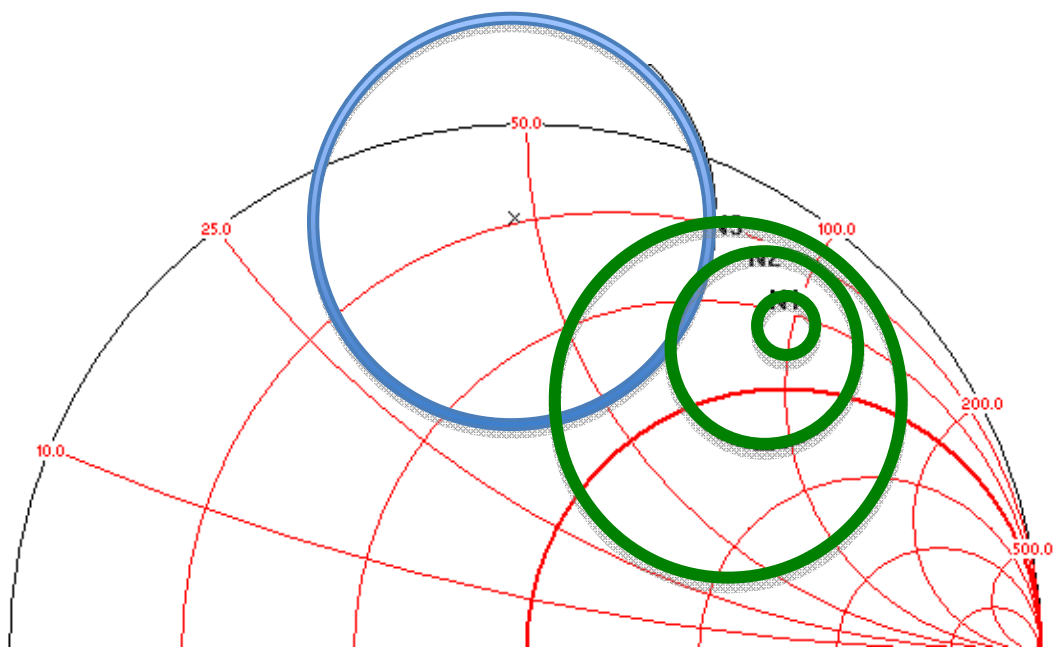

Figure 3. Noise (green) and gain (blue) circles

From the 2-port network theory, described by scattering parameters, the reflection coefficient at Port 2 of a device whose Port 1 is terminated in a load with a reflection coefficient $\Gamma_s$ is given by:

$$\Gamma_{OUT} = S_{22} + \frac{S_{12}S_{21}\Gamma_S}{1 - S_{11}\Gamma_S} \quad (4)$$

Using (4) $\Gamma_{OUT}$ was calculated and the corresponding impedance was 53+j84 Ω. The output matching network must present an impedance $Z_L$=53-j84 Ω at the transistor output for maximum power transfer. The next step is to design the matching networks considering the source and load impedances calculated above. The main challenge is to design lossless matching networks that will provide the lowest NF for a wide bandwidth. For all the devices the matching networks type is presented in Fig. 4.

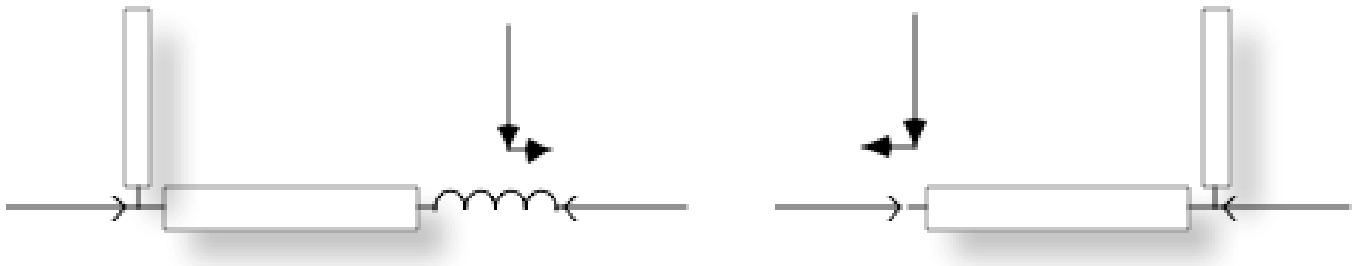

Figure 4. Input and output maching networks

The simulations were performed using the S and Noise parameters provided by the manufacturers and the required transmission lines were designed for the RT5880 (Rogers Corporation) substrate.

An additional 3 Ω resistor in series with signal output line was included to improve the circuit stability [7]. The final circuit is represented in Fig. 5.

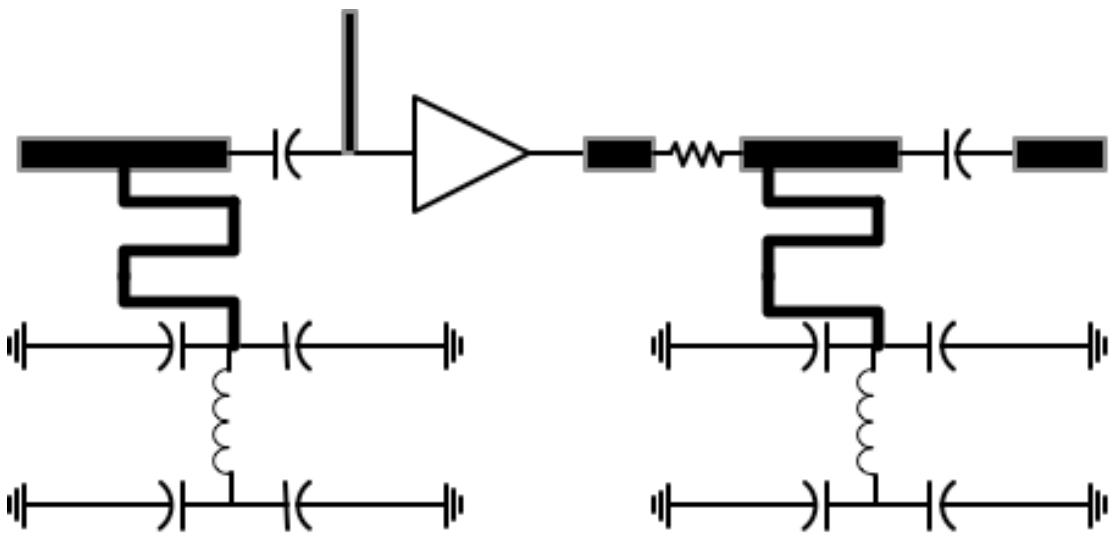

Figure 5. Final circuit used in simulation

The microstrip lines dimensions and nominal values of the lumped elements were optimized by trial and error in the simulations. The best results for NF are summarized in TABLE III.

TABLE III. SIMULATION RESULTS OF ALL PHEMTS

| DEVICE | NF (dB) @ 5 GHZ | GAIN (dB) |
|---|---|---|
| MGF4941 | 0.25 | 13.8 |
| MGFC4419 | 0.28 | 12.3 |
| MGF4931 | 0.48 | 11.0 |
| ATF-36077 | 0.32 | 13.0 |
| NE3511S02 | 0.25 | 13.0 |
| FHX13X | ----- | 14.0 |

## IV. PROTOTYPE IMPLEMENTATION AND TESTING RESULTS

The simulation results are presented in Fig. 6. The results obtained in TABLE III are good enough for our purpose.

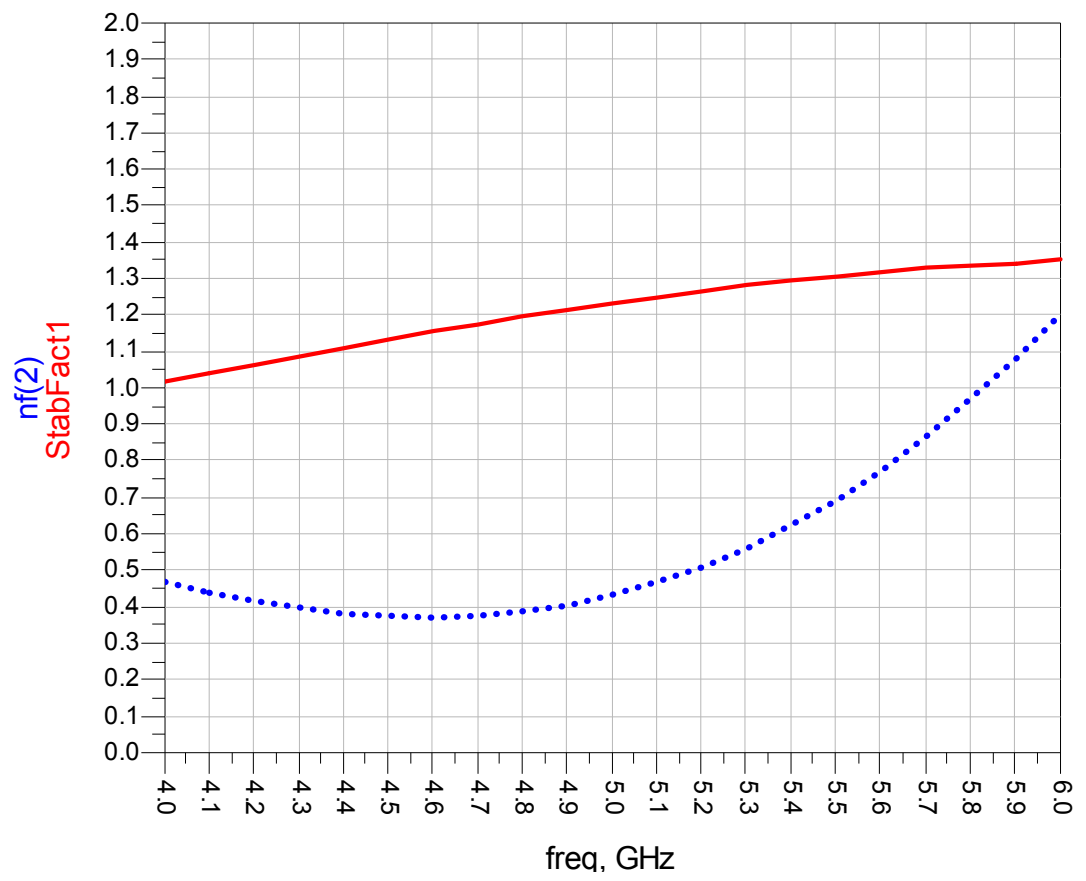

Figure 6. Simulation: Noise Figure and Stability Factor (K)

The final layout (Fig. 7) was designed in AutoCAD and a positive photographic film was created for PCB manufacturing.

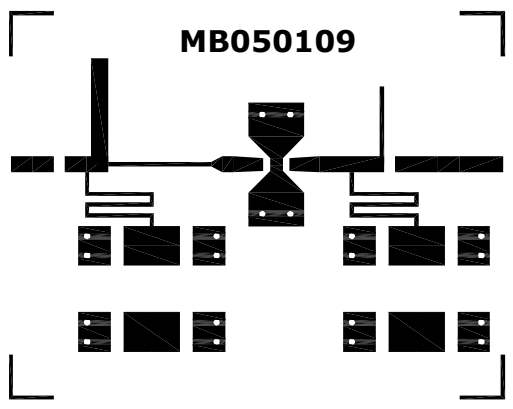

Figure 7. Final Layout for PCB fabrication

The handmade RF inductor chokes, used for polarization decoupling, were built using a thin wire wrapped forming an air coil and placed between the two capacitors pairs (Fig. 5). Thick film type resistors were used and the high characteristic impedance (very narrow) microstrip line at the pHEMT gate behaves as the small matching inductance. The capacitors in the signal lines are DC blocks. The PCB was placed in an aluminum-milled box and tested as shown in Fig. 8.

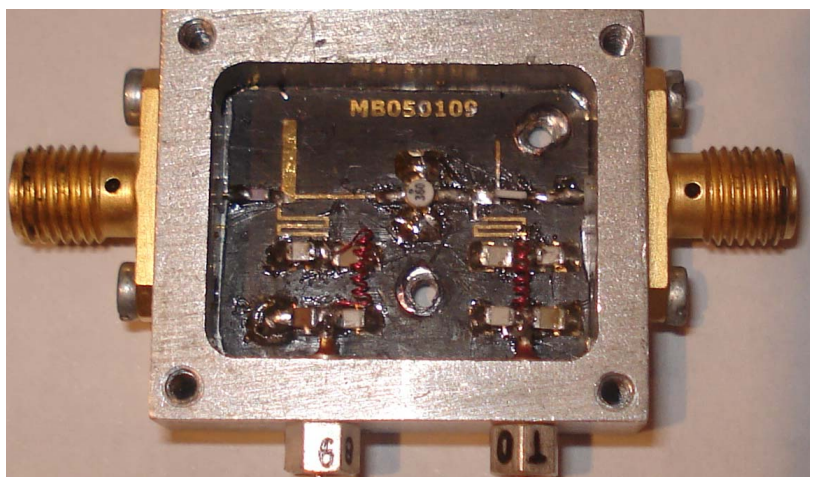

Figure 8. The prototype prepared for testing

### A. *Gain Measurement*

The gain was measured with an HP8753E Network Analyzer. A comparison of the simulated and measured results is presented in figure (Fig. 9). In the desired operating frequency the measured gain is only about 1 dB less than simulated.

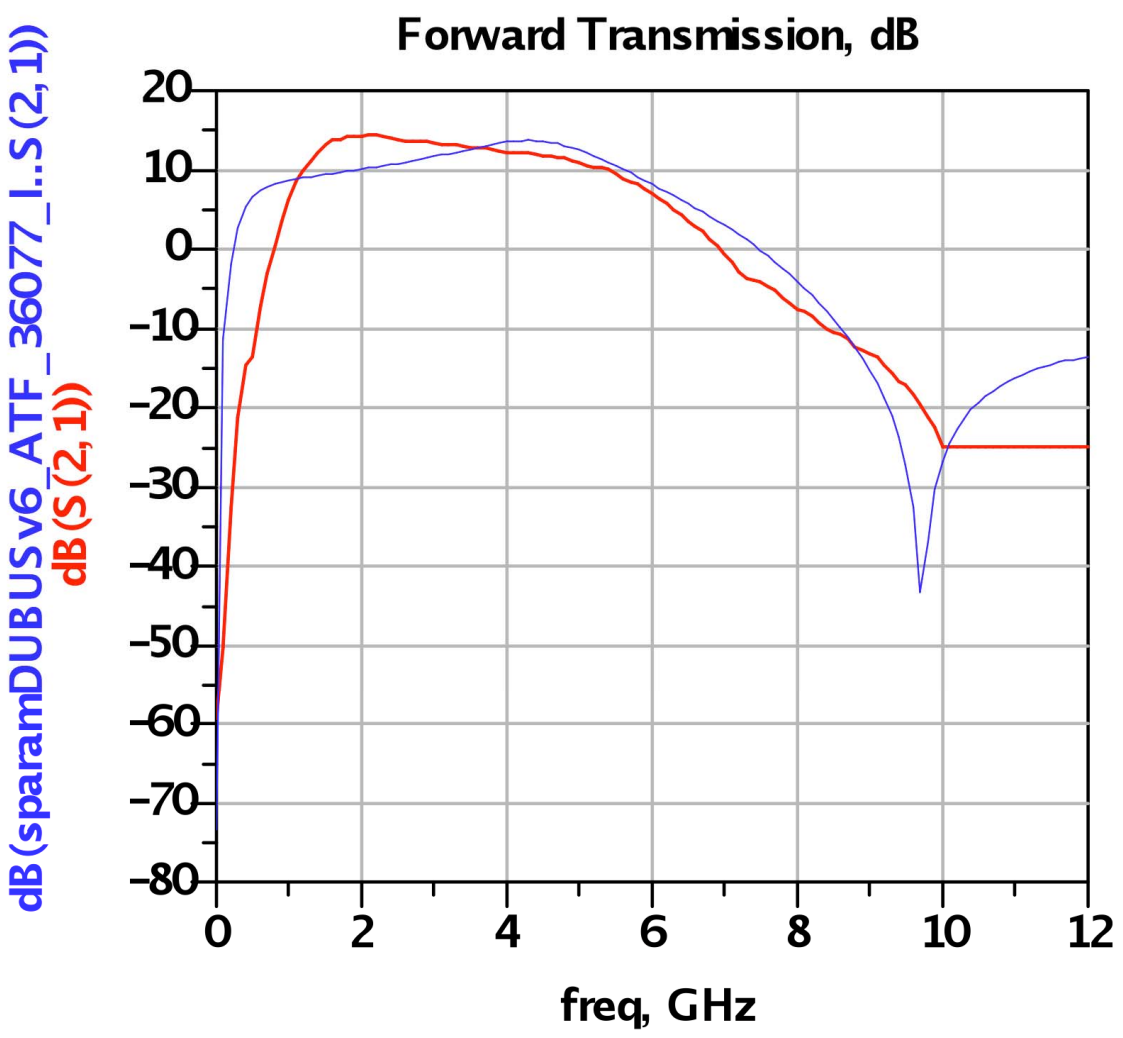

Figure 9. Gain values: simulated (blue) and measured (red)

### B. *Noise Figure Measurements*

The measurement setup arrangement [8,9] is described in Fig. 10 and the components used were:

- NF meter (HP8970A)
- Noise Head HP346C/ ENR (calibration +/-0.02 on Nov 2009)
- Attenuator (7 dB) NARDA type 4872-7 7.00dB +/- 0.01dB (RL <-22 dB)
- Amplifier (AFT6232-Avantek 4.5 dB NF or MITEC 108660 3 dB NF)
- Mixer (M2 0115 – Marki)
- Isolator (REC 3A6NH)

plus cables and connectors.

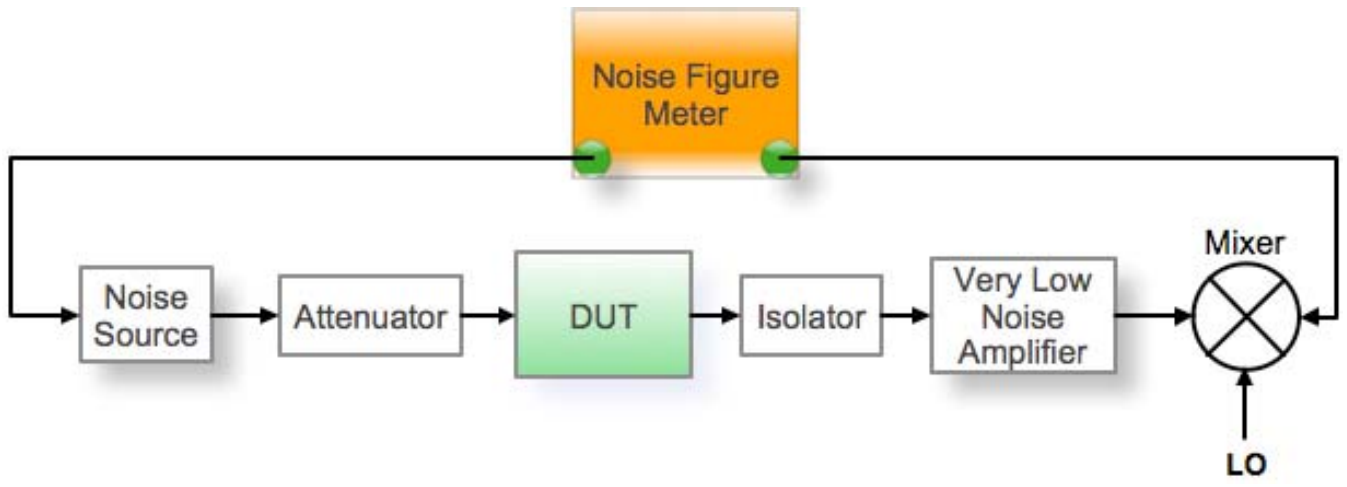

Figure 10. Test Bench used in Noise Figure Measurement

The NF was measured at the target application frequencies: 4.8, 4.9 and 5 GHz. Several NF measurements were made at each frequency using different bias voltages.

The best noise performance was obtained with the gate and drain voltages $V_{GS}$ = 0,9 V and $V_{DS}$ = 1,6 V. It must be pointed out the measurement accuracy of +/- 0.01 dB. The final results are summarized in TABLE IV.

TABLE IV. NOISE FIGURE MEASUREMENTS RESULTS

| FREQUENCY (GHz) | NF (dB) |
|---|---|
| 4.8 | 0.62 |
| 4.9 | 0.61 |
| 5.0 | 0.62 |

The experimental results show about 0.2dB difference to simulation. Several factors may be contributing to this difference namely the scattering parameters for the best biasing conditions may differ slightly from the used ones; losses in the used ceramic multilayer SMD capacitors and radiation losses.

The capacitors and resistors package were too large for the microstrip line widths (W =0,7 mm), so some rearrangements on the layout or new lumped components must be used in new designs.

## V. CONCLUSIONS

A market survey of low noise devices to be used in LNA amplifier design operating at room temperatures was performed. Several LNAs were simulated. An affordable prototype was developed and tested giving an excellent noise performance at room temperature (Noise temperature of about 45 K). It may probably still be further optimized. Due to its low cost the present LNA can be a solution for mass production targeting Radio astronomy applications where many amplifiers are required.

## ACKNOWLEDGMENTS

The author acknowledge PANORAMA for PhD support. Barbosa is supported by a Ciência 2007 Fellowship, under QREN and COMPETE funding. The team acknowledges support of GEM-P-project through FCT contract PTDC/CTE-AST/65925/2006. We thank José Carlos Pedro for useful discussions and comments.